\documentclass[a4paper,11pt]{article}
\pdfoutput=1 

\usepackage{jheppub} 

\usepackage[T1]{fontenc} 
\usepackage{multirow}

\title{\boldmath Scalar triplet on a domain wall: an exact solution}

\author[a,b,1]{Vakhid A. Gani,\note{Corresponding author.}}
\author[b,c]{Mariya A. Lizunova,}
\author[d]{Roman V. Radomskiy}

\affiliation[a]{Department of Mathematics, National Research Nuclear University MEPhI\\ (Moscow Engineering Physics Institute), 115409 Moscow, Russia}
\affiliation[b]{Theory Department, National Research Center Kurchatov Institute, Institute for Theoretical and Experimental Physics, 117218 Moscow, Russia}
\affiliation[c]{Department of Theoretical Nuclear Physics, National Research Nuclear University MEPhI\\ (Moscow Engineering Physics Institute), 115409 Moscow, Russia}
\affiliation[d]{Department of Elementary Particle Physics, National Research Nuclear University MEPhI\\ (Moscow Engineering Physics Institute), 115409 Moscow, Russia}

\emailAdd{vagani@mephi.ru}

\abstract{
We study a model with a real scalar Higgs field and a scalar triplet field that allows existence of a topological defect --- a domain wall. The wall breaks the global $O(3)$ symmetry of the model, which gives rise to non-Abelian orientational degrees of freedom. We found an exact analytic solution that describes a domain wall with a localized configuration of the triplet field on it. This solution enables one to calculate contributions to the action from the orientational and translational degrees of freedom of the triplet field. We also study the linear stability of the domain wall with the triplet field switched off. We obtain that degrees of freedom localized on the wall can appear or do not appear depending on the parameters of the model.
}

\begin{document}

\maketitle

\flushbottom

\section{\label{sec:level1} Introduction}

Topological defects in field models are of great interest for modern theoretical physics from cosmology to condensed matter. The subject is extremely wide, therefore we direct the reader to modern reviews, e.g., \cite{vilenkin,manton}. We remark the
impressive progress in scenarios with embedded topological defects, e.g., a skyrmion on a domain wall, or a Q-lump on a domain wall~\cite{nitta1,nitta2,nitta3,nitta4,jennings,nitta5,blyankinshtein}. There are also interesting results on topological and non-topological solitons~\cite{luchini01}, multisoliton configurations~\cite{saadatmand,gumerov}, and on interaction of solitary waves with one another~\cite{dorey01,GaKuPRE,oliveira01,oliveira02,GaKuLi,GaLeLi,GaLeLiconf,Ahlqvist:2014uha,Mohammadi} and with impurities~\cite{saad01,saad02,saad03,krusch01}. Various modifications of the Skyrme model have also been actively studied recently~\cite{krusch02,krusch03,nitta6}.

In this work we discuss partial breaking of a non-Abelian symmetry due to the presence of a topological defect (in particular, a domain wall).
Non-Abelian degrees of freedom have been widely discussed in the recent literature, in particular, in connection with non-Abelian strings~\cite{hanany00,auzzi,shifman00,hanany,lilley,shifman01,shifman02,shifman03,shifman04,gorsky,nitta} that can arise both in supersymmetric and non-supersymmetric models, see also reviews~\cite{shiyu,konishi,tong}.

Consider a field model with a global non-Abelian symmetry $G$ that enables the existence of topological defects (strings, vortices, domain walls). These defects break the symmetry to its subgroup $H$, also global. The topological defect is then parameterized, in addition to the translational modes, by $\dim G - \dim H$ orientational degrees of freedom. These orientational modes are sigma-fields in the $G/H$ coset.

The authors of the recent Ref.~\cite{kur} have studied a simple model with a real scalar Higgs field $\varphi$ and a scalar triplet field $\chi^i$ that allows existence of a domain wall that breaks the global symmetry of the model. As a result, in addition to the translational degree of freedom, two orientational modes appear. 
Ref.~\cite{kur} also introduced the spin-orbit interaction that leads to entanglement between the translational and rotational degrees of freedom. Most of certainly interesting and important results of Ref.~\cite{kur} have been obtained numerically.

We consider the field model suggested in Ref.~\cite{kur}, particularly, we study configurations that can be described as ``domain wall + a lump of the field $\chi$''. Using a functional Ansatz for the dependence of the field $\varphi$ on the coordinate perpendicular to the domain wall, we obtain an exact analytic solution of the equations of motion. From this solution, we find contributions to the action that correspond to the orientational degrees of freedom. We also study the linear stability of the ``bare'' domain wall, i.e., a solution in the form of ``domain wall + $\chi\equiv 0$'', and derive analytic expressions for the excitation spectrum.
We also obtain constraints on the parameters of the model that follow from the
requirement that the ``bare'' domain wall be stable.

Our paper is organized as follows. In Section~\ref{sec:level2} we introduce the model and show how one can obtain an exact analytical solution using a particular Ansatz. We also discuss some of the main properties of the obtained solution and calculate the terms in the action that describe the translational and orientational degrees of freedom. We add the spin-orbit interaction in that section too. In Section~\ref{sec:level5} we study the linear stability of the ``bare'' domain wall. We conclude with a discussion of the results and the prospects for future research in Section~\ref{sec:level6}. In the Appendix we compare our units with those used in Ref.~\cite{kur}, and notice a discrepancy between our value of the lowest excitation
frequency and that reported in Ref.~\cite{kur}.

\section{\label{sec:level2} Domain wall}

\subsection{The model}

\begin{figure}
\begin{center}
	\includegraphics[scale=0.5]{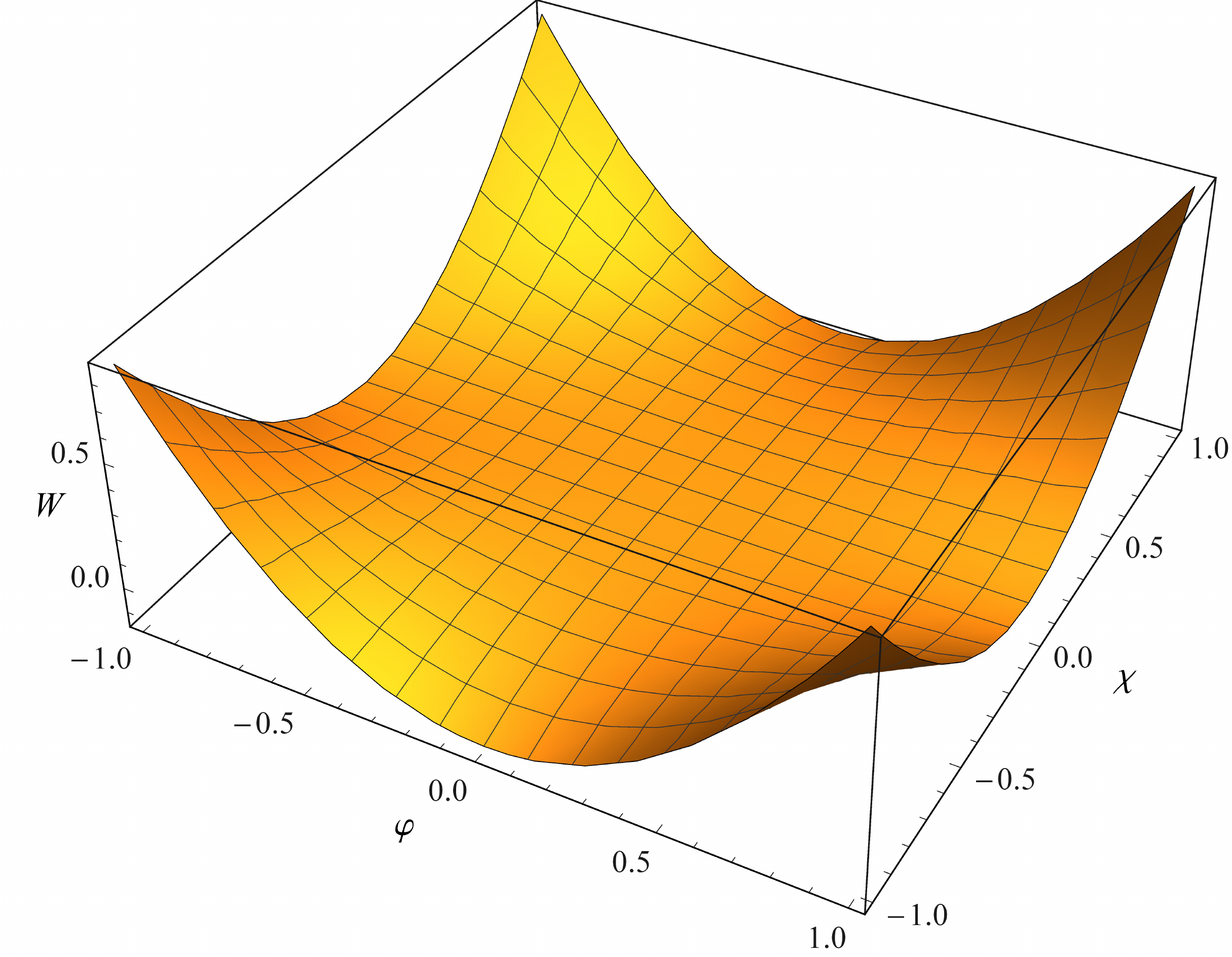}
\end{center}
	\caption{Potential \eqref{eq:Potential_W} as a function of fields $\varphi$ and $\chi$, for $\beta=0.1$, $\mu=0.2$. Note that the range of $\chi$ is extended to negative values.}
	\label{fig:pot_w}
\end{figure}

We consider a real scalar field $\varphi$ and a triplet of real scalars $\chi^i$ coupled to $\varphi$ in (3+1)-dimensional space-time. The dynamics of the system is determined by the Lagrangian
\begin{equation}\label{eq:Lagrangian}
\mathcal{L}=\frac{1}{2}\partial_{\nu}\varphi\partial^{\nu}\varphi+\frac{1}{2}\partial_{\nu}\chi^i\partial^{\nu}\chi^i-V(\varphi)-\gamma W(\varphi,\chi)
\end{equation}
with the potentials being
\begin{equation}\label{eq:Potential_V}
V(\varphi)=(\varphi^2-1)^2,
\end{equation}
\begin{equation}\label{eq:Potential_W}
W(\varphi,\chi^i)=(\varphi^2-\mu^2)\chi^i\chi^i+\beta(\chi^i\chi^i)^2,
\end{equation}
where $\gamma,\beta\ge 0$, $0 \le \mu^2\le 1$ are constants, and the index $i$ runs from  1 to 3. At $\gamma=0$ this Lagrangian describes a Higgs-like field $\varphi$ and a free scalar triplet $\chi^i$ that do not interact with each other.
The interaction potential $W(\varphi,\chi^i)$ depends only on $\chi^2=\chi^i\chi^i$ --- there is no dependence on the orientation of $\chi^i$ in the internal space.
In Fig.~\ref{fig:pot_w} we show the potential~\eqref{eq:Potential_W} for a particular
set of values of the model parameters.

The Lagrangian yields the following equations of motion:
\begin{equation}\label{eq:eqmo_d}
\begin{cases}
  \varphi_{tt}-\Delta\varphi=-4(\varphi^2-1)\varphi-2\gamma\varphi\chi^2,\\
   \chi_{tt}^i-\Delta\chi^i=-2\gamma(\varphi^2-\mu^2)\chi^i-4\beta\gamma\chi^2\chi^i.
\end{cases}
\end{equation}
We are interested in static configurations that resemble planar domain walls. Assuming the wall to be perpendicular to the $z$ axis,
\begin{equation}\label{eq:case}
\varphi=\varphi(z), \quad \chi^i=\chi^i(z),
\end{equation}
and choosing a particular orientation of $\chi^i$,
\begin{equation}\label{eq:orient_chi}
\chi^i(z)=\chi(z)\left(
\begin{array}{c}
0\\
0\\
1\\
\end{array}
\right),
\end{equation}
the equations of motion for $\varphi(z)$ and $\chi(z)$ become
\begin{equation}\label{eq:eqmo_main}
  \begin{cases}
   \varphi^{\prime\prime}=4(\varphi^2-1)\varphi+2\gamma\varphi\chi^2,\\

\chi^{\prime\prime}=2\gamma(\varphi^2-\mu^2)\chi+4\beta\gamma\chi^3,
 \end{cases}
\end{equation}
where the prime stands for the derivative with respect to $z$.

\subsection{Exact solution}

To find an exact analytic solution of~\eqref{eq:eqmo_main}, we first consider simple heuristic arguments that follow from the Lagrangian~\eqref{eq:Lagrangian}: if $\gamma=0$, the field $\chi^i$ is free, and the dynamics of $\varphi$ is described by the potential of the $\varphi^4$ model,
 with the kink
\begin{equation}\label{eq:kink}
\varphi_{\mbox {\scriptsize w}} (z)=\tanh\left(\sqrt{2}z\right)
\end{equation}
being the solution we look for. We further assume that when the interaction between the fields is turned on, $\varphi(z)$ will keep the shape of a kink~\eqref{eq:kink}, possibly with a different spatial scale~\cite{lensky,GaLiRaconf}, i.e., we look for the solutions of the system~\eqref{eq:eqmo_main} using the Ansatz
\begin{equation}\label{eq:phi}
\varphi_{\mbox {\scriptsize s}}(z)=\tanh\alpha z.
\end{equation}
Inserting this in the first equation of the system~\eqref{eq:eqmo_main} gives
\begin{equation}\label{eq:chi}
\chi_{\mbox {\scriptsize s}}(z)=\frac{A}{\cosh\alpha z},
\end{equation}
where $A$ is a constant given by
\begin{equation}\label{eq:A0}
A^2=\frac{2-\alpha^2}{\gamma}.
\end{equation}
Eqs.~\eqref{eq:phi} and \eqref{eq:chi}, plugged in the second equation of the system \eqref{eq:eqmo_main}, give two more relations between the constants:
\begin{equation}\label{eq:alpha}
\alpha^2=2\gamma q,
\quad
\alpha^2=\gamma(1-2\beta A^2),
\end{equation}
here $q=1-\mu^2$. Finally, inserting the first of equalities \eqref{eq:alpha} in \eqref{eq:A0}, we obtain
\begin{equation}\label{eq:A}
A^2=\frac{2(1-\gamma q)}{\gamma}.
\end{equation}

Eqs.~\eqref{eq:phi} and \eqref{eq:chi}, together with the constraints on the constants, thus give a solution of the equations of motion. From \eqref{eq:alpha} and \eqref{eq:A} follows a relation between the model parameters that has to be fulfilled so as
to allow the existence of the solution found above:
\begin{equation}\label{eq:cond1}
\gamma=\frac{4\beta}{2q(2\beta-1)+1}.
\end{equation}
As $\alpha^2\ge 0$ and $A^2\ge 0$ we get:
\begin{equation}\label{eq:cond2}
\frac{4\beta}{1+4\beta q} \le \gamma\le \frac{1}{q}.
\end{equation}

Let us now briefly examine some of the properties of the solution that we obtained.
\begin{figure}
\begin{center}
	\includegraphics[scale=0.7]{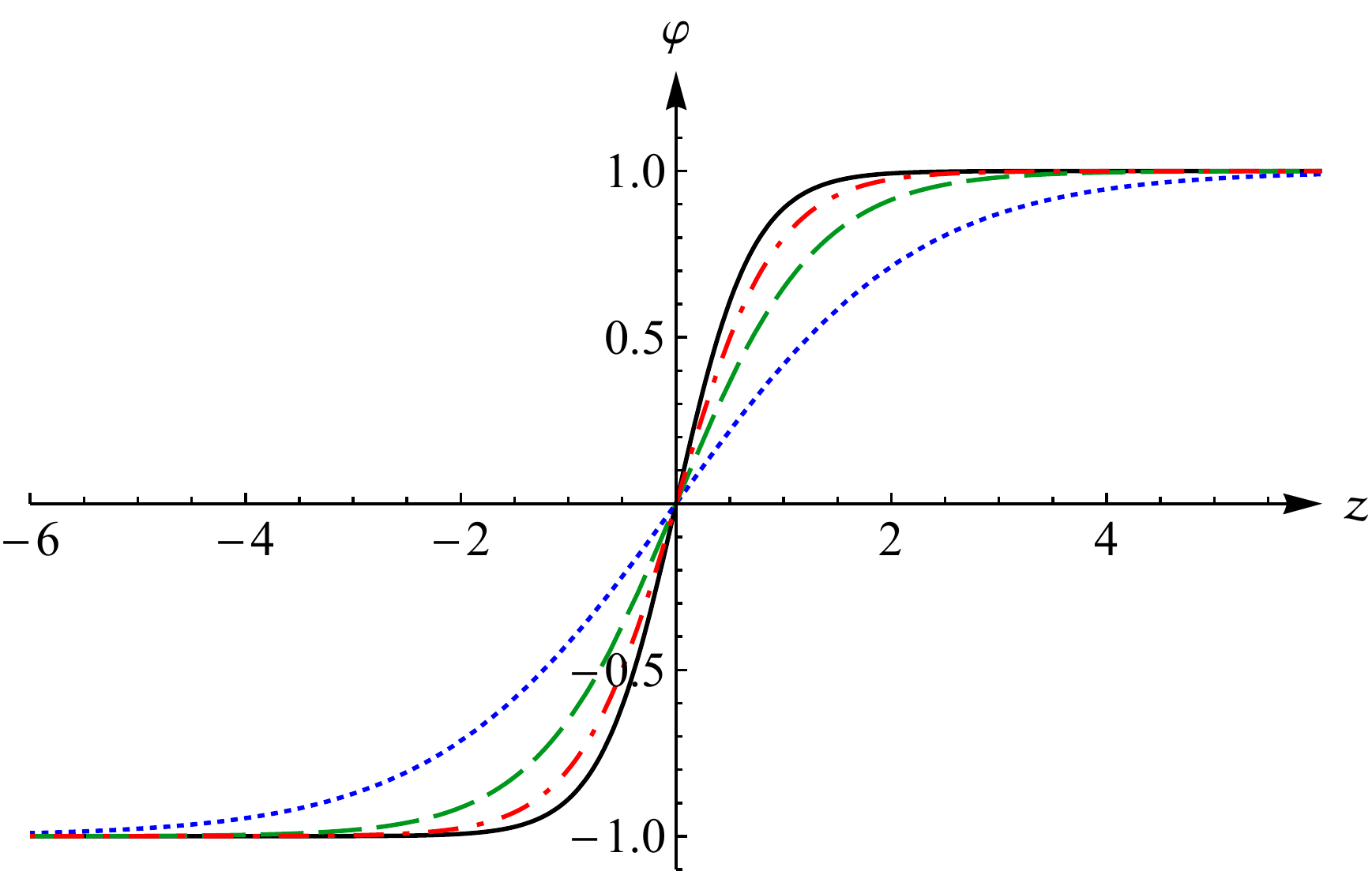}
\end{center}
	\caption{Function $\varphi_{\mbox {\scriptsize s}}(z)$ for $\gamma=1$ and three different values of $q$: $0.1$ (blue dotted line), $0.3$ (green dashed line), and $0.6$ (red dash-dotted line). The solid line shows $\varphi=\varphi_{\mbox {\scriptsize w}} (z)$.}
	\label{fig:phi}
\end{figure}
\begin{figure}
\begin{center}
	\includegraphics[scale=0.7]{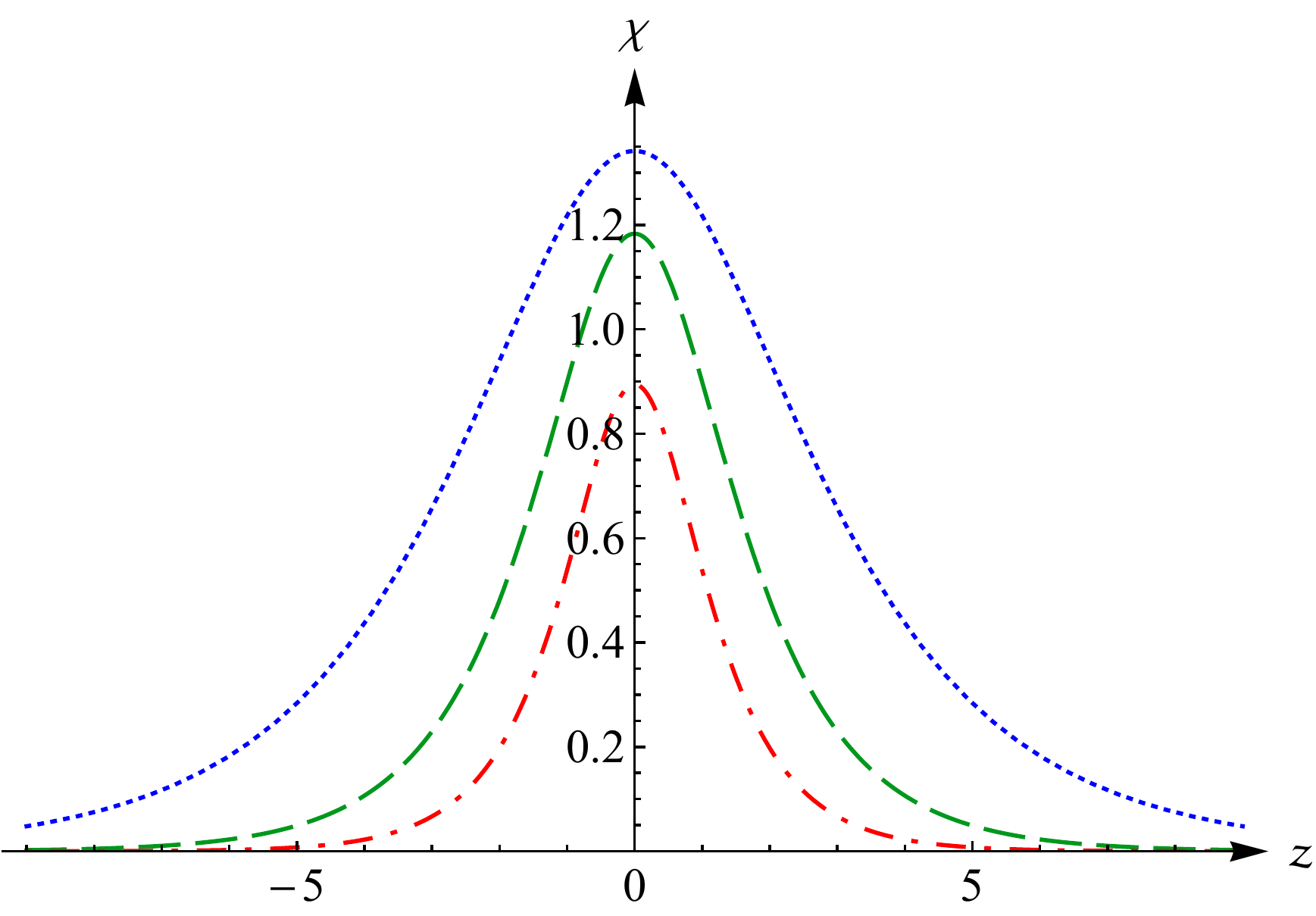}
\end{center}
	\caption{Function $\chi_{\mbox {\scriptsize s}}(z)$ for $\gamma=1$ for the same values of $q$ as in Fig.~\ref{fig:phi}.}
	\label{fig:chi}
\end{figure}
\begin{figure}
\begin{center}
	\includegraphics[scale=0.7]{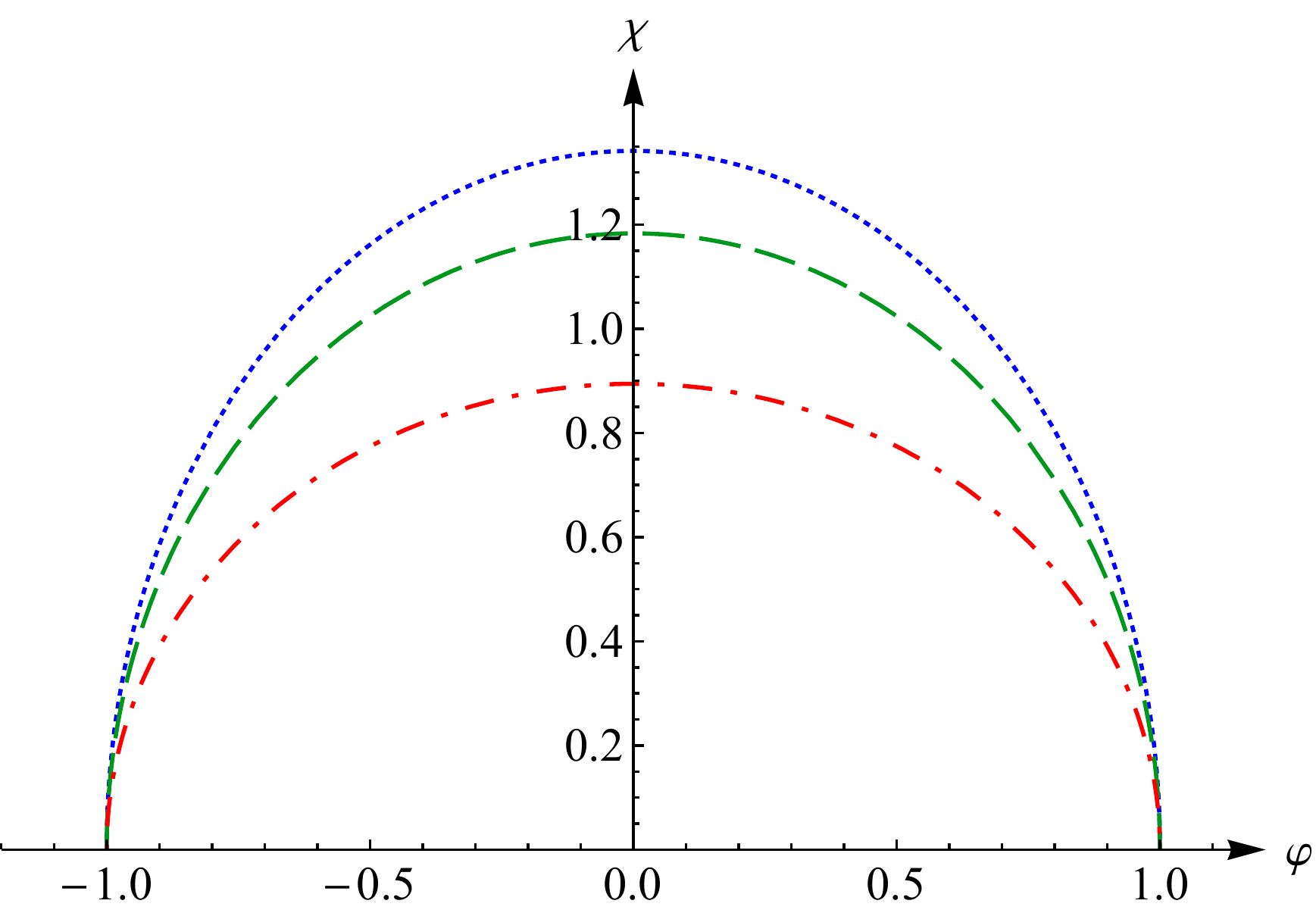}
\end{center}
	\caption{Trajectories \eqref{eq:ell} in $(\varphi,\chi)$ plane for $\gamma=1$ for the same values of $q$ as in Fig.~\ref{fig:phi}.}
	\label{fig:chi_phi}
\end{figure}
Firstly, it follows from \eqref{eq:A0} that $A=0$ at $\alpha=\sqrt{2}$, and we have the domain wall \eqref{eq:kink} with $\chi(z)\equiv 0$. Secondly, the scale over which $\chi$ is significantly different from its vacuum value(s) is of the order of $\alpha^{-1}$, which coincides with the same scale of the field $\varphi$, see Figs.~\ref{fig:phi} and \ref{fig:chi}. Thirdly, the asymptotics of the fields at $z\to-\infty$ are
\begin{equation}
\varphi_{\mbox {\scriptsize s}}\approx-1+2\:e^{2\alpha z}, \quad \chi_{\mbox {\scriptsize s}}\approx 2A\:e^{\alpha z},
\end{equation}
coinciding with those reported in Ref.~\cite{kur}.
The asymptotics at $z\to+\infty$ are the following:
\begin{equation}
\varphi_{\mbox {\scriptsize s}}\approx 1-2\:e^{-2\alpha z}, \quad \chi_{\mbox {\scriptsize s}}\approx 2A\:e^{-\alpha z}.
\end{equation}
Fourthly, notice that the solution \eqref{eq:phi}, \eqref{eq:chi} parameterizes a semi-ellipse in the plane $(\varphi,\chi)$ (assume $A>0$):
\begin{equation}\label{eq:ell}
\frac{\chi^2}{A^2}+\varphi^2=1.
\end{equation}
This ellipse degenerates into the segment $-1\le\varphi\le 1$, $\chi=0$ at $\alpha=\sqrt{2}$, see Fig.~\ref{fig:chi_phi}.

Finally, the energy of the solution \eqref{eq:phi}, \eqref{eq:chi}
is given by
\begin{equation}\label{eq:energyeq}
\begin{split}
E_{\mbox{\scriptsize s}}&=\int\limits_{-\infty}^{+\infty}\left[\frac{1}{2}\left(\frac{d\varphi}{dz}\right)^2+\frac{1}{2}\left(\frac{d\chi}{dz}\right)^2+(\varphi^2-1)^2+\gamma\left[(\varphi^2-\mu^2)\chi^2+\beta\chi^4\right]\right]dz\\
&=\frac{4A^4\beta\gamma+2(\alpha^2+2)+A^2(\alpha^2+2\gamma(3q-2))}{3\alpha}.
\end{split}
\end{equation}

\subsection{Moduli}

Using our solution, we can find analytic expressions for the terms in the action on the wall world-volume that correspond to the orientational and translational moduli \cite{shifman01,shifman02}. We define the following integrals appearing in the derivation
below:
\begin{equation}\nonumber
\begin{split}
J_1&=\int\chi_{\mbox {\scriptsize s}}^2(z)dz=2\:\sqrt{\frac{2}{\gamma q}}\:\frac{1-\gamma q}{\gamma},\\
J_2&=\int\left(\frac{d\chi_{\mbox {\scriptsize s}}}{dz}\right)^2dz=\frac{2}{3}\:\sqrt{\frac{2}{\gamma q}}\:\frac{1-\gamma q}{\gamma},\\
J_3&=\int\left(\frac{d\varphi_{\mbox {\scriptsize s}}}{dz}\right)^2dz=\frac{4\sqrt{2\gamma q}}{3} .
\end{split}
\end{equation}
As the orientation of $\chi^i$ is arbitrary, we can write
\begin{equation}
\chi^i=\chi_{\mbox {\scriptsize s}}(z)P^i(t,x,y),
\end{equation}
where $P^i$ is a unit vector in the internal space. The density of the kinetic term
of the field $\chi$ in the Lagrangian is then rewritten as
\begin{equation}
\partial_\mu\chi^i\partial^\mu\chi^i=\chi^2_{\mbox {\scriptsize s}}(z)\partial_\nu P^i\partial^\nu P^i-\left(\frac{d\chi_{\mbox {\scriptsize s}}}{dz}\right)^2,\quad \nu=0,1,2.
\end{equation}
Integrating over $z$ we arrive to the corresponding contribution to the action:
\begin{equation}
\Delta S_1=\frac{J_1}{2}\int\partial_\nu P^i\partial^\nu P^i dt\:dx\:dy.
\end{equation}

The domain wall breaks the translational symmetry of the physical space
along the $z$ direction. Parameterizing the shifts of the field $\chi$ by $\chi_{\mbox {\scriptsize s}}(z-z_0(t,x,y))$, we obtain the corresponding kinetic term in the form:
\begin{equation}
\left(\frac{d\chi_{\mbox {\scriptsize s}}}{dz}\right)^2(\partial_\nu z_0\partial^\nu z_0-1),\quad \nu=0,1,2.
\end{equation}
Integration over $z$ gives the following contribution to the action:
\begin{equation}
\Delta S_2=\frac{J_2}{2}\int(\partial_\nu z_0\partial^\nu z_0)\:dt\:dx\:dy.
\end{equation}
Similarly for the field $\varphi$ we have
\begin{equation}
\Delta S_3=\frac{J_3}{2}\int(\partial_\nu z_0\partial^\nu z_0)\:dt\:dx\:dy.
\end{equation}
The solution \eqref{eq:phi}, \eqref{eq:chi} that we found can
be also used in order to calculate the contribution of the 
spin-orbit interaction to the action, as done numerically in Ref.~\cite{kur}.
The spin-orbit interaction is introduced by adding the term
$-\varepsilon(\partial_i\chi^i)^2$ to the Lagrangian, see \cite{shifman01} for detail.
Here, $\varepsilon$ is a small parameter; our solution \eqref{eq:phi}, \eqref{eq:chi}
yields the following spin-orbit term in the action, at the leading
order in the expansion over powers of $\varepsilon$:
\begin{equation}
\begin{split}
\Delta S_4=&-\varepsilon\: J_2\int\left[(\partial_k z_0)(\partial_l z_0)P^kP^l+P^3P^3+2(\partial_k z_0)P^kP^3\right]dt\:dx\:dy\\
&-\varepsilon\: J_1\int\left(\partial_k P^k\right)^2dt\:dx\:dy,\quad k,l=1,2.
\end{split}
\end{equation}

\section{\label{sec:level5} Instability of the domain wall with $\mathbf{\boldsymbol{\chi}\equiv 0}$}

We consider small perturbations $\varphi(t,z)$ and $\chi(t,z)$ around the static solution \eqref{eq:case}-\eqref{eq:orient_chi} in the following form \cite{bazeia01,bazeia02}:
\begin{equation}
\varphi(t,z)=\varphi_0(z)+\delta\varphi(t,z),\quad
\chi(t,z)=\chi_0(z)+\delta\chi(t,z).
\end{equation}
Linearizing \eqref{eq:eqmo_d} with respect to $\delta\varphi$ and $\delta\chi$ results in
\begin{equation}
 \begin{cases}
   \delta\varphi_{tt}-\delta\varphi_{zz}=-12\varphi_0^2\delta\varphi+4\delta\varphi-2\gamma\chi_0^2\delta\varphi-4\gamma\chi_0\varphi_0\delta\chi,\\
   \delta\chi_{tt}-\delta\chi_{zz}=-2\gamma\varphi_0^2\delta\chi+2\gamma(1-q)\delta\chi-12\beta\gamma\chi_0^2\delta\chi-4\gamma\chi_0\varphi_0\delta\varphi.
 \end{cases}
\end{equation}
We use the following standard Ansatz for the perturbations:
\begin{equation}
\delta\varphi(t,z)=e^{-i\omega t}\phi(z),\quad \delta\chi(t,z)=e^{-i\omega t}\zeta(z).
\end{equation}
It results in the following spectral problem:
\begin{equation}
\hat{H}
\left(\begin{array}{crl}
\phi\\
\zeta
\end{array}\right)=\omega^2 \left(\begin{array}{crl}
\phi\\
\zeta
\end{array}\right),
\end{equation}
where
\begin{equation}
\hat{H}= \left(\begin{array}{crl}
-\displaystyle\frac{d^2}{dz^2}+12\varphi_0^2+2\gamma\chi_0^2-4&4\gamma\chi_0\varphi_0\\
4\gamma\chi_0\varphi_0&-\displaystyle\frac{d^2}{dz^2}+2\gamma\varphi_0^2+12\beta\gamma\chi_0^2-2\gamma(1-q)
\end{array}\right).
\end{equation}

Using this scheme we now investigate the stability of the ``bare'' domain wall \eqref{eq:kink} with $\chi\equiv 0$, i.e.~$\varphi_0=\varphi_{\mbox{\scriptsize w}}(z)$, $\chi_0\equiv 0$. In this case the operator $\hat{H}$ becomes
\begin{equation}
\hat{H}= \left(\begin{array}{crl}
-\displaystyle\frac{d^2}{dz^2}+12\varphi_{\mbox{\scriptsize w}}^2-4&0\\
0&-\displaystyle\frac{d^2}{dz^2}+2\gamma\varphi_{\mbox{\scriptsize w}}^2-2\gamma(1-q)
\end{array}\right).
\end{equation}
The equations for $\phi$ and $\zeta$ decouple \cite{bazeia02}:
\begin{equation}\label{eq:unlink_phi}
-\phi_{zz}+4(3\varphi_{\mbox{\scriptsize w}}^2-1)\phi=\omega^2\phi,
\end{equation}
\begin{equation}\label{eq:unlink_chi}
-\zeta_{zz}+2\gamma(\varphi_{\mbox{\scriptsize w}}^2-1+q)\zeta=\omega^2\zeta.
\end{equation}
At this point one can use the well-known result for the modified P\"oschl-Teller potential. Eq.~\eqref{eq:unlink_phi} can be written in the form
\begin{equation}
\phi_{zz}+2\left(\frac{\omega^2}{2}-4+\frac{6}{\cosh^2(\sqrt{2}z)}\right)\phi=0.
\end{equation}
The discrete spectrum of this equation has two levels:
\begin{equation}
\omega_n^2=2n(4-n),\quad n=0, 1,
\end{equation}
which gives $\omega_0^2=0$ and $\omega_1^2=6$. This result does not depend on the parameters of the model, as can be deduced already from Eq.~\eqref{eq:unlink_phi}. The spectrum of the perturbations of the field $\chi$ is given by Eq.~\eqref{eq:unlink_chi} that, in turn, can be rewritten in the form
\begin{equation}
\zeta_{zz}+2\left(\frac{\omega^2}{2}-\gamma q+\frac{\gamma}{\cosh^2(\sqrt{2}z)}\right)\zeta=0.
\end{equation}
The number of levels in the discrete spectrum is bound by $n<s=\left(\sqrt{1+4\gamma}-1\right)/2$. The frequency eigenvalues are
\begin{equation}
\omega_n^2=-2\gamma-1+2\gamma q-2n(1+n)+(1+2n)\sqrt{1+4\gamma}.
\end{equation}
The lowest discrete level corresponds to $n=0$:
\begin{equation}\label{eq:om_zero}
\omega_0^2=-2\gamma+2\gamma q+\sqrt{1+4\gamma}-1.
\end{equation}
The stability against small perturbations demands that $\omega_0^2\ge 0$,
which results in the following constraint on the model parameters: 
\begin{equation}\label{eq:cond_omega0ge0}
\omega_0^2\ge 0 \quad \Longrightarrow \quad \gamma\le \frac{q}{(1-q)^2}.
\end{equation}
This inequality defines the range of the model parameters where the ``bare'' domain wall \eqref{eq:kink} with $\chi\equiv 0$ is linearly stable, and the field $\chi^i$ does not appear on the wall.

If the parameters do not satisfy the inequality \eqref{eq:cond_omega0ge0}, the domain wall \eqref{eq:kink} is unstable with respect to small perturbations in linear approximation. This means that initially formed configuration \eqref{eq:kink} with $\chi\equiv 0$ will evolve into the configuration of the type of \eqref{eq:phi}, \eqref{eq:chi}.

\section{\label{sec:level6} Discussion}

In the present study we have obtained an analytic solution of the considered field
model. This solution consists of a domain wall with a localized configuration of the scalar triplet field on the wall. This solution is similar
to that studied in Ref.~\cite{kur} numerically.
We derive conditions that have to be satisfied by the parameters of the model
in order for the solution that we found to exist. It is interesting that
no such condition has been identified in Ref.~\cite{kur}.

Even though our exact solution \eqref{eq:phi}, \eqref{eq:chi} is similar to
the numerical solution found in Ref.~\cite{kur},
the model parameters used in that work, inserted in our solution, result in $A^2<0$.
We, however, restrict our study to the model sectors that satisfy \eqref{eq:cond1},
\eqref{eq:cond2}. Using our solution, we obtain the terms in the action that describe
the translational and orientational degrees of freedom.

We also study the linear stability of the ``bare'' domain wall \eqref{eq:kink} with $\chi\equiv 0$, analytically obtaining the spectrum of small excitations.
Our results confirm that this is unstable at the values of the model parameters
used in Ref.~\cite{kur}. We also notice a discrepancy between our value of the lowest
excitation frequency and that reported in Ref.~\cite{kur}, see the Appendix.
We also derive conditions under which the ``bare'' domain wall is linearly stable.

In conclusion, we emphasize that, depending on the parameters of the model, degrees of freedom localized on the wall can appear or do not appear.

This work opens wide prospects for further research. In particular, it would be useful to trace the relation between our exact solution and the numerical solution of Ref.~\cite{kur}. Besides that, a study of the interaction of two parallel domain walls similar to those considered in this work would be of interest; the collective coordinate method could be useful in this context, see, e.g., \cite{GaKuLi,luchini02}.

\section*{Acknowledgments}

The authors are very grateful to Dr.~V.~Lensky for critical comments that resulted in substantial improvement of the manuscript. V.~A.~Gani acknowledges the support of the Ministry of Education and Science of the Russian Federation, Project No.~3.472.2014/K. M.~A.~Lizunova thanks the ITEP support grant for junior researchers and gratefully acknowledges financial support from the Dynasty Foundation.

This work was performed within the framework of the Center of Fundamental Research and Particle Physics supported by the MEPhI Academic Excellence Project (contract No.~02.a03.21.0005, 27.08.2013).

\section*{\label{sec:level7} Appendix. Comparison of our units with those used in Ref.~\cite{kur}}

The Lagrangian of the model studied in Ref.~\cite{kur} is
\begin{equation}\label{eq:lagrangian_old}
\widetilde{\mathcal{L}}=\frac{1}{2}\tilde{\partial}_{\nu}\tilde{\varphi}\tilde{\partial}^{\nu}\tilde{\varphi}+\frac{1}{2}\tilde{\partial}_{\nu}\tilde{\chi}^i\tilde{\partial}^{\nu}\tilde{\chi}^i-\tilde{\lambda}(\tilde{\varphi}^2-\tilde{v}^2)^2-\tilde{\gamma}\left[(\tilde{\varphi}^2-\tilde{\mu}^2)\tilde{\chi}^i\tilde{\chi}^i+\tilde{\beta}(\tilde{\chi}^i\tilde{\chi}^i)^2\right],\quad \tilde{v}^2>\tilde{\mu}^2,
\end{equation}
which is easy to show to be equivalent to our system \eqref{eq:Lagrangian}.
Namely, passing to dimensionless units, we obtain
\begin{equation}
\mathcal{L}=\frac{1}{2}\partial_\nu\varphi\partial^\nu\varphi+\frac{1}{2}\partial_\nu\chi^i\partial^\nu\chi^i-(\varphi-1)^2-\gamma\left[(\varphi^2-\mu^2)\chi^i\chi^i+\beta(\chi^i\chi^i)^2\right],
\end{equation}
where
\begin{equation}
\widetilde{\mathcal{L}}/(\tilde{\lambda}\tilde{v}^4)=\mathcal{L},\quad \tilde{\varphi}/\tilde{v}=\varphi,\quad \tilde{\chi}^i/\tilde{v}=\chi^i,\quad \tilde{x}^\nu\tilde{v}\sqrt{\tilde{\lambda}}=x^\nu,\quad \tilde{\gamma}/\tilde{\lambda}=\gamma,\quad \tilde{\mu}/\tilde{v}=\mu,\quad\tilde{\beta}=\beta\gamma.
\end{equation}
This coincides with the Lagrangian used in this work, Eq.~\eqref{eq:Lagrangian}. The  parameters of the Lagrangian \eqref{eq:lagrangian_old} were set in Ref.~\cite{kur}
to
\begin{equation}\label{eq:constant_value_o}
\tilde{\lambda}=\frac{1}{12},\quad\tilde{\beta}=0.2,\quad \tilde{\gamma}=\frac{2}{3},\quad\frac{\tilde{\mu}}{\tilde{v}}=0.55,
\end{equation}
which translates to
\begin{equation}\label{eq:constant_value_n}
\beta=0.2,\quad\gamma=8,\quad \mu=0.55,
\end{equation}
in the dimensionless units used in our work.

The authors of Ref.~\cite{kur} perform a numerical study of the linear stability
of the ``bare'' domain wall, using the values given by \eqref{eq:constant_value_o}.
These values, inserted into our condition \eqref{eq:cond_omega0ge0}, yield $\gamma\le 7.62$, which indeed is not satisfied, as $\gamma=8$.
Therefore at least the lowest frequency $\omega_0^2$ is negative, and the configuration is unstable. We found that at the above values of the model parameters the excitation spectrum contains three discrete levels:
\begin{equation}
\omega_0^2\simeq -0.0954,\quad \omega_1^2 \simeq 7.394,\quad\omega_2^2\simeq 10.883.
\end{equation}
Our value of $\omega_0^2$ differs from the value that follows from the study of Ref.~\cite{kur}.

\end{document}